\documentclass[letter]{aa}
\usepackage{txfonts}
\usepackage{graphicx}
\usepackage{amsmath}
\usepackage{color}
\usepackage{gensymb}

\bibpunct{(}{)}{;}{a}{}{,} 

\usepackage{hyperref}

\newcommand{\equ}[1]{Eq.~(\ref{#1})}
\newcommand{\fig}[1]{Fig.~\ref{#1}}
\newcommand{\Fig}[1]{Figure~\ref{#1}}

\newcommand{\equs}[2]{Eqs.~(\ref{#1})-(\ref{#2})}
\newcommand{\Equ}[1]{Equation~(\ref{#1})}

\renewcommand{\vec}[1]{\protect\mathbf{#1}}
\newcommand{\grad}{\vec{\boldsymbol{\nabla}}}
\newcommand{\curl}{\vec{\boldsymbol{\nabla}}\mathbf{\times}}
\renewcommand{\div}{\vec{\boldsymbol{\nabla}}\cdot}
\newcommand{\f}[2]{\frac{#1}{#2}}
\newcommand{\dpart}[2]{\f{\partial #1}{\partial #2}}

\newcommand{\vA}{\vec{A}}
\newcommand{\vU}{\vec{u}}
\newcommand{\vB}{\vec{B}}

\newcommand{\vJ}{\vec{J}}
\newcommand{\vE}{\vec{E}}

\newcommand{\vEMF}{\boldsymbol{\mathcal{E}}}
\newcommand{\mean}[1]{\overline{#1}}
\newcommand{\fluct}[1]{#1}

\newcommand{\fluctB}{\fluct{b}}

\newcommand{\fluctvA}{\fluct{\vec{a}}}
\newcommand{\fluctvB}{\fluct{\vec{b}}}
\newcommand{\fluctvU}{\fluct{\vec{u}}}

\newcommand{\meanB}{\mean{B}}

\newcommand{\meanvA}{\mean{\vA}}
\newcommand{\meanvB}{\mean{\vB}}

\newcommand{\meanvEMF}{\mean{\vEMF}}

\begin{document}
\title{Ultimate large-$Rm$ regime of the solar dynamo}
\titlerunning{Ultimate large-$Rm$ regime of the solar dynamo}
\author{F. Rincon\inst{1,2}}
\institute{CNRS; IRAP; 14 avenue Edouard Belin, F-31400 Toulouse,
           France,
\and 
Universit\'e de Toulouse; UPS-OMP; IRAP: Toulouse, France\\
\email{frincon@irap.omp.eu}}

\date{\today}
\abstract{For more than 40 years the quest to understand how large-scale magnetic fields emerge from turbulent flows in rotating astrophysical systems, such as the Sun, has been a major focus of computational astrophysics research.
Using a parameter scan and phenomenological analysis of maximally simplified  three-dimensional cartesian magnetohydrodynamic simulations of 
large-scale non-linear  helical turbulent dynamos, I present results in this Letter that strongly point to an asymptotic ultimate regime of 
the large-scale solar dynamo at large magnetic Reynolds numbers, $Rm$, involving helicity fluxes between hemispheres. I obtained
corresponding numerical solutions at both $Pm>1$ and $Pm<1$, and show that they can currently only be achieved in clean, simplified numerical set-ups.  
The analysis further strongly suggests that all global simulations to date lie in non-asymptotic turbulent magnetohydrodynamic 
 regimes highly 
sensitive to changes in kinetic and magnetic Reynolds numbers. Ideas are presented to attempt to reach the ultimate regime 
in such 'realistic' global spherical models at a reasonable numerical cost.
Overall, the results clarify the current 
state, and some hard limitations of the brute-force numerical modelling approach applied to this, and other similar astrophysical turbulence problems.
}

\keywords{Sun: magnetic fields -- Dynamo -- Turbulence -- Magnetohydrodynamics (MHD)}
\maketitle

\section{Introduction}
\begin{figure*}
\centering\includegraphics[width=\textwidth]{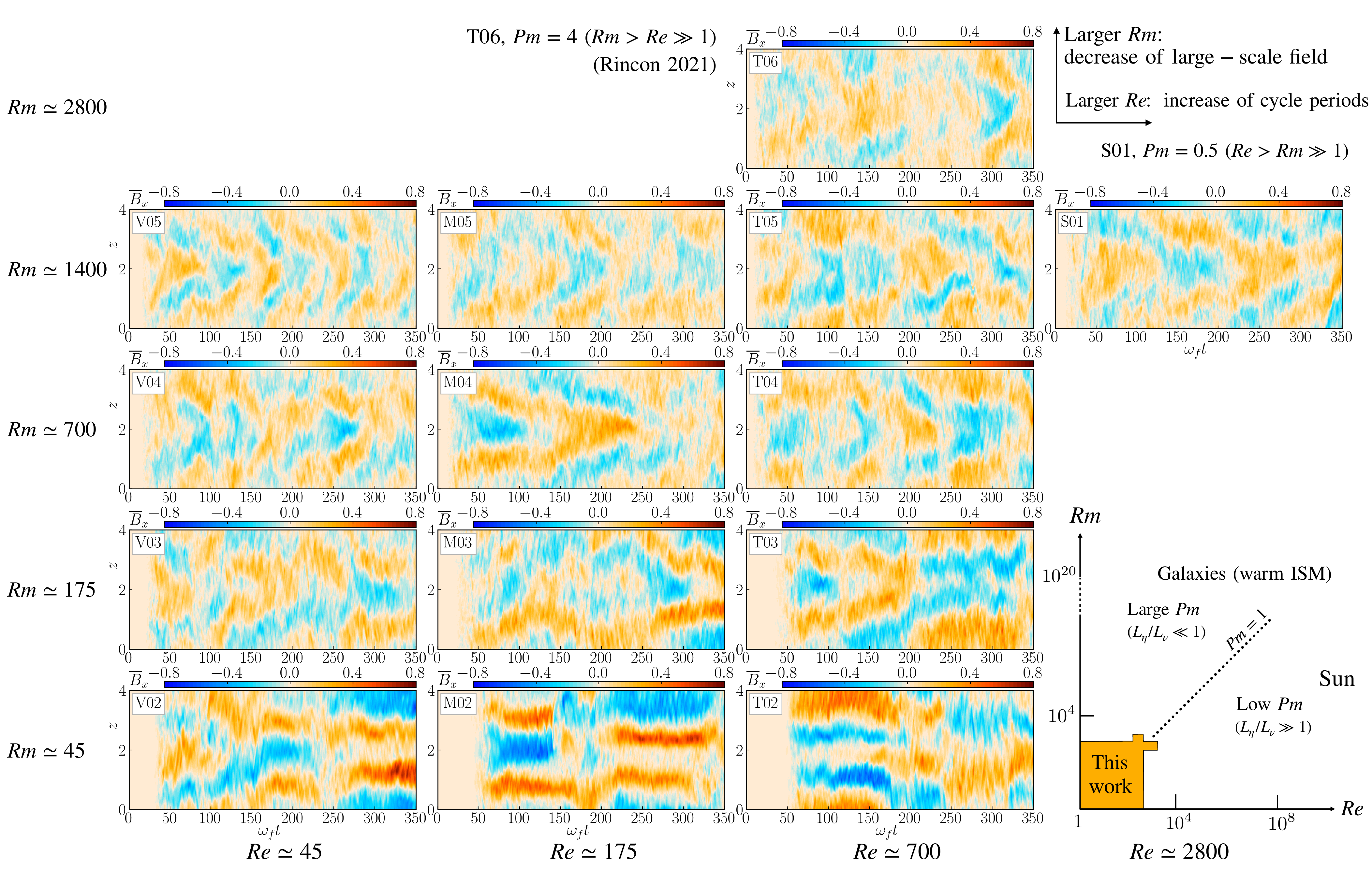}
\caption{Butterfly diagrams $\meanB_{x}(z,t)$ of large-scale non-linear helical dynamo modes vs $Re$ and $Rm$.
Run S01 ($Pm=0.5$) is the rightmost plot; run T06 ($Pm=4$) of R21 is the topmost plot. The parameter range spanned
is shown at the bottom right. The same colour scale is used in all plots.\label{fig1}}
\end{figure*}

The making of large-scale magnetism by turbulent fluid flow in symmetry-broken rotating systems, 
such as stars or galaxies, is known as the large-scale dynamo effect, and is an archetypal case of highly
multiscale turbulent dynamics in astrophysics. This problem presents us with major theoretical challenges,
starting with its intrinsically three-dimensional nature; we have 
come to rely increasingly on high-performance computing (HPC) to decipher these challenges.
Regular numerical progress has been achieved in modelling the convectively driven solar dynamo
\citep{parker55}
since the pioneering work of \citet{gilman83} (see reviews by \citealt{branden05,rincon19,charbonneau20,kapyla25} 
and, over the last ten years, modelling work 
by \citealt{guerrero16,hotta16,kapyla17,strugarek18,hotta21,brun22,kapyla23,warnecke25}).
Magnetohydrodynamic (MHD)  modelling of the solar cycle nevertheless remains a numerical quagmire; there are discrepancies between models currently
attributed to several physical effects, including rotation, magnetic feedbacks, geometry, and to specific 
numerical implementations \citep{charbonneau20}. However,  an even more general question looms: turbulence regimes 
accessible to simulations, notably characterised by the kinetic and magnetic Reynolds 
numbers, $Re$ and $Rm$, remain far  from the astrophysical realm (bottom right of \fig{fig1}). Accordingly, 
we need to understand how far we are from accurate computer models of astrophysical dynamos.
Are they even on the horizon ?

Addressing these questions requires  proceeding methodically towards large $Re$ and $Rm$,
preferably with the right ordering between the two. However, global models, many relying on numerical dissipation, 
have favoured large-scale realism (spherical geometry, differential rotation, radial stratification) 
at the expense of turbulent non-linearities and explicit dissipation. While tricks can be used to probe 
linear large-scale dynamos at large $Rm$  \citep{tobias13}, in minimal controlled set-ups with 
reduced effective dimensionality, 
they cannot describe the relevant dynamical non-linear regimes \citep{pongkiti16}.  Overall, limited work 
has been devoted (in solar physics at least; see e.g. \cite{sheyko16,schaeffer17,aubert17} for the faster-rotating geodynamo)
to a careful exploration via direct numerical simulation (DNS), using explicit viscosity and resistivity,  of the 
non-linear regime at large $Re$ and $Rm$. This is an issue for several reasons; for instance, the large-$Rm$ asymptotics of 
catastrophic quenching of helical dynamos \citep{branden05} cannot be studied with global simulations \citep{simard16}. 
\cite{delsordo13}, and more recently \cite{rincon21} (hereafter R21) and \cite{branden25}, have attempted to explore in simpler 
local cartesian set-ups how non-linear dynamos with hemispheric helicity distributions, such as those expected from rotating convection,
change with $Re$ and $Rm$ and, in relation to catastrophic quenching, what dominant magnetic-helicity budget balances are satisfied asymptotically. In addition,
most models have magnetic Prandtl numbers $Pm>1$, opposite to the Sun. 
Overall, many uncertainties remain as to how small- and large-scale dynamics couple at large $Rm$ \citep{warnecke25}. 

This Letter expands significantly on R21 to probe large-scale non-linear helical dynamos at large $Re$ and $Rm$. 
I present new analyses that point to an ultimate large-$Rm$ solar dynamo regime
involving magnetic helicity fluxes, and compute such numerical solutions at both $Pm>1$ and $Pm<1$. 
A standardised comparison with global simulations suggests that they are not in this regime,
and I pinpoint their key intrinsic limitations. I also discuss the current limitations 
of my own results, and possible ways for global models to reach the ultimate regime at reasonable cost. 

\section{Cartesian helical dynamo simulations}
I considered the 3D MHD numerical experiment introduced in R21 (see Appendix~\ref{appequations})
to study the non-linear phase of a helically driven large-scale turbulent dynamo effect. 
In a spatially periodic, cartesian box elongated along the $z$-direction, a turbulent flow was forced
at the $(x,y)$ box scale by a Galloway-Proctor-inspired helical forcing term reversing at $z=0$
\citep{galloway92}.
Anticipating the large-scale nature of the emerging dynamo mode, 
the box was made larger by a factor of $4$ in the $z$ direction than in the $(x,y)$ directions, $L_z/(L_x,L_y)=4$ in order to allow a minimal scale-separation 
between the turbulence injection scale, $L_f\sim L_x$, and the scale $\sim L_z$ of the dynamo mode itself.  
This forcing and set-up ensures a  hemispheric  sinusoidal distribution of kinetic helicity 
reversing at the `equator' (here $z=0$, $L_z/2$, see \fig{fig2}), typical of rotating turbulent astrophysical 
systems, such as solar convection banana cells.
The goal, then, is to grasp how the 
non-linear dynamo properties change with $Re=u_\mathrm{rms}/(k_f\nu)$ and $Rm=u_\mathrm{rms}/(k_f\eta)$. Here, $k_f=2\pi/L_f$ 
is the forcing wavenumber, $u_\mathrm{rms}$ is the rms turbulent velocity, $\nu$ is the kinematic 
viscosity, and $\eta$ is the magnetic diffusivity. 

This maximally simplified setting still captures the physical essence of what I think is minimally required to drive a spatially distributed
large-scale dynamo in a rotating turbulent system, such as the Sun or a galaxy. The experiment is fully 3D and non-linear,
and has built-in scale-separation without the burden of
complex geometry and stratification effects. Its turbulence has a realistic order-one correlation to turnover time ratio, with a well-defined 
hemispheric turbulent helicity distribution, but zero net volume-averaged helicity, a key feature to escape the catastrophic 
dynamo quenching (see e.g. \citealt{branden01} and R21). This simplicity enables the use of exponentially fast converging 
spectral methods to probe large $Re$ and $Rm$ regimes with an optimal use of the numerical resolution 
and, critically, all dissipative processes under control. Using the incompressible MHD spectral code 
SNOOPY with 2/3 dealiasing \citep{lesur07} at resolutions up to $512^2\times 2048$, one can reach large 
$Re,Rm=O(3000)$ by astrophysical MHD standards. 

The R21 simulations are complemented by a new high-resolution run, labelled S01, with 
$Re=2800$ and $Pm=\nu/\eta=0.5$ (Table~\ref{tab1}), to probe the $Pm<1$ behaviour of the dynamo, 
typical of the solar interior. Instead, the highest-resolution T06 simulation in R21 at $Rm=2800$ was  focused on pushing into the
large-$Rm$ asymptotics, and limited to $Pm=4$ to adequately resolve all scales. Each simulation was run for a 
minimum of 50 forcing times $2\pi/\omega_f$ (up to $\sim 200$ actual flow turnover times $L_f/u_\mathrm{rms}$) to allow 
for the statistical emergence, saturation, and long-time evolution of a large-scale dynamo mode.
A mosaic overview of butterfly diagrams of the $x$-component $\meanB_{x}(z,t)$ of the $(x,y)$-averaged magnetic field 
emerging in the simulations is shown in Fig.~\ref{fig1}. In all runs, a large-scale mode is excited and dynamically sustained. 
Two prominent trends can be seen:
\begin{itemize}
\item As $Rm$ ($Pm$) increases at constant $Re$, the system bifurcates from a bistable-like steady state to a migrating wave state, 
and the magnitude of $\meanB$ decreases (see Table~\ref{tab1} and \fig{fit};
\smallskip).
\item At large $Rm$, the dynamo wave period increases with increasing $Re$ (decreasing $Pm$); at lower $Rm$, the system also transitions from a steady state to a wave as $Re$ increases.
\end{itemize}

The results of the new S01 run $(Pm=0.5)$ are similar to the larger $Rm$, lower $Re$ run T06 $(Pm=4)$. This 
includes the dynamo wave pattern and large-scale field saturation level (Table~\ref{tab1}). 
Their time-averaged kinetic and magnetic energy spectra are shown in Fig.~\ref{fig3}.
The magnetic and kinetic dissipative cutoffs, $k_\eta=2\pi/L_\eta$ and $k_\nu=2\pi/L_\nu$ 
are  closer in S01, as expected. However, the factor of eight difference in $Pm$ only has a modest effect 
on the overall mode and turbulent cascades.

Magnetic helicity is an important near-invariant in this problem \citep{branden05,moffatt16,kleeorin22},
and a detailed analysis of its dynamics is key to unlocking the complexity of this landscape. 
Building on R21, I show in Appendix~\ref{regimesection} that the simulations can be divided into three distinct 
regimes (\fig{fig4}): a low-$Rm$ resistive ($R$) regime up to $Rm\sim 50$; an intermediate ($I$) regime up to $Rm=O(500)$,
still subject to a strong small-scale magnetic-helicity quenching bottleneck; and an asymptotic large-$Rm$ ultimate ($U$)
regime where resistive quenching of helicity is asymptotically subdominant at both large and small scales. 
There,
the dominant balance in both large- and small-scale helicity budgets is between electromotive-force-driven helicity generation 
and the divergences of the $z$-oriented mean fluxes of helicity, two processes independent of resistivity that non-linearly 
adjust to each other during the self-consistent dynamical evolution. 

This phase diagram illuminates \fig{fig1}.
At low $Rm$, each hemisphere develops its own steady catastrophically quenched helical dynamo, and there is little 
communication between the two hemispheres except when $Re$, and therefore turbulent diffusion becomes significant, thereby triggering weak
helicity fluxes in $z$. The qualitative change in the nature of the solutions at $Rm>50$ is therefore interpreted as a symptom of the transition 
between the low and intermediate regimes. In the ultimate large-$Rm$ regime, on the other hand, magnetic helicity fluxes dominate over
resistive dissipation of helicity even at low $Re$, and are thus freely exchanged between the two hemispheres. This allows the dynamo to
escape catastrophic quenching and gives rise to hemispheric synchronicity in the form of migrating $\alpha^2$-dynamo waves, 
whose period depends on the strength of turbulent diffusion, controlled by $Re$. Of all runs, only those with $Rm>1000$ lie comfortably 
in the ultimate regime. This includes T05, T06 at $Pm\geq 1$, and, importantly, also the new run S01 at $Pm<1$.

In Appendix~\ref{diffusive}, I  show that a satisfactory diffusive transport model can be devised to fit, and also interpret
the simulation results. This model recovers all three regimes with reasonable parameters, such as turbulent magnetic diffusion, fractional helicities, and effective fluctuation and mean-field scales,
fitted to the simulation data (\fig{diffusivefit}). Most 
importantly, it correctly predicts the decrease in $\meanB$ with $Rm$ in simulations, including the rather sluggish 
asymptote into the ultimate regime (\fig{fit}).

\section{Comparison with global models}
Table~\ref{tabcompare} provides a comparison of the (standardised) parameter regimes of recent spherical global models of the 
solar dynamo, and of idealised cartesian models of large-scale helical dynamo with spatially reversing helicity distributions: 
$Re$, $Rm$ (as defined in this work on forcing wavenumbers), $Pm$, and $k_\mathrm{f}/\overline{k}$ (Appendix~\ref{diffusive}; 
when $k_f$ was not provided, $k_f=\ell_\mathrm{peak}/R_\odot$, where $\ell_\mathrm{peak}$ is the peak harmonic of the kinetic 
energy spectrum, and $\overline{k}=1-2/R_\odot$, corresponding to a dipolar or quadrupolar field, were used; *$Pm=1.46$ 
from \citep{hotta16} was used to estimate $Rm$ in \cite{hotta21}).
Local DNS models fare much better in terms of $Re$ and $Rm$, both due to their spectral 
convergence (for R21 and this work) and to smaller controlled scale separations between 
the turbulent injection scale and the box scale, which provides more resolution for
turbulent dissipative structures. 
Both \fig{fig1} and the analyses in Appendices~\ref{regimesection}-\ref{diffusive} suggest that most global models are far from the regime 
touched by T06 and S01. Most lie somewhere in the lower left intermediate-regime quarter of \fig{fig1}, where the dynamo pattern appears most 
sensitive to $Re$ and $Rm$.  This may go a long way towards explaining why the outcomes of global simulations, including cycle periods,
are extremely model-dependent \citep{charbonneau20}, and vary significantly as $Pm$ barely 
changes at mild $Rm$ \citep[e.g.][]{kapyla17}.  Our diffusive model, in particular \equs{bbar1}{bbar3} for $\meanB$, further 
suggests that different fractional helicities injected in convection at different Rossby numbers (encapsulated by the $\theta$ 
parameters of the model) can significantly contribute to the scatter and rotational dependence of global models.

\setlength{\tabcolsep}{5pt}
\begin{table}[h]
\caption{Parameters of earlier global and local non-linear simulations.\label{tabcompare}}
\centering\begin{tabular}{lcccr}
\hline\hline
Study (run) & $Re$ & $Rm$ & $Pm$ & $k_\mathrm{f}/\overline{k}$ \\ 
\hline
\cite{strugarek18} (O5) & 54 & 73 & 1.35 & 10-20 \\ 
\cite{kapyla17} (G3) & 134 & 134 & 1 & 10-20 \\ 
\cite{hotta16} (H) & 313 & 457 & 1.46 & 10-20 \\ 
\cite{warnecke25} (4M) & 550 & 550 & 1 & 10-20 \\ 
\cite{hotta21} (H)
 & 1130 & 1650* & 1.46* & 15-30 \\ 
\hline
\citeauthor{branden25} (P)
& 1700 & 340 & 0.2 & 4-8 \\
\cite{delsordo13} (S6) & 1063 & 1063 & 1 & 4-8 \\
R21, this work (T06) & 695 & 2778 & 4 & 4-8 \\ 
This work (S01) & 2904 & 1452 & 0.5 & 4-8 \\ 
\hline\hline
\end{tabular}
\end{table}

The H-model of \cite{hotta21} is currently the only global model approaching the turbulent regimes of the  
T06 and S01 runs. This can be seen by comparing the spectra in their Fig.~4 with \fig{fig3} 
here. In all cases, the ratio of the peak turbulence injection scale to the magnetic dissipation scale is $\sim 50$. 
However, there is a subtle but key caveat here. As pointed out by \cite{mitra10b}, and made explicit in Appendix~\ref{diffusive}, 
the critical $Rm_{I-U}$ separating the intermediate and ultimate regimes has a strong dependence $\propto (k_f/\overline{k})^2$ on the scale-separation 
between the mean field and turbulent injection scales. In the H-run of \cite{hotta21}, the spectral peak of convection is shifted 
towards rather small scales (spherical harmonics $\ell\gtrsim 30$) compared to a large-scale spherical dipole or quadrupole,
making this scale-ratio ($k_f/\overline{k}\sim 15-30$) much higher than in the present controlled set-up ($k_f/\overline{k}\sim 4-8$; see 
Table~\ref{tabcompare} and Appendix~\ref{diffusive}). Because of this large-scale separation inherent to global models 
(and likely to the Sun itself), I estimate that achieving the ultimate regime  with realistic global models may require $Rm>5000$. 
Accordingly, and despite their massive resolution and similar $Rm$ to the present highest-resolution runs, the runs of \cite{hotta21} likely lie 
in the core of the intermediate regime, not in the ultimate regime as for S01 and T06. This is further supported by their report (Fig.~3b)
of a decrease in $\meanB$ with increasing $Rm$, typical of the intermediate regime (\equ{bbar2} and \fig{fit}). 
It is also notable in this respect that they find no true oscillating large-scale field in their high-resolution run.

The effective $Pm$ of most global models so far is larger than one, which may be a problem with respect to solar realism. 
Here, the new results hint at a little piece of good news: the overall similarity between the $Pm=0.5$ and $Pm=4$ runs 
suggests that large-scale dynamos with hemispheric helicity distributions, and their saturation 
at large $Rm$, may only be weakly dependent on $Pm$ at large $Re$.
Further support for this hypothesis comes from my earlier remark that turbulent helicity fluxes, 
which are key to the excitation of dynamo waves, should plateau at large $Re$.

Finally, S01 is above the critical small-scale dynamo (SSD) threshold
at $Pm=0.5$;  considering its similarity with T06 at $Pm=4$, this suggests that the issue of SSD--no SSD 
\citep{hotta16,hotta21,warnecke25} may only be secondary to the production of a dynamo with a large-scale component 
(Kazantsev-model analyses \citep{malyshkin07,malyshkin09} suggest that the helical dynamo at large $Rm$ is a self-consistent 
unified multiscale mode like that I simulated, not the mere composition of different modes). The relative vigor of small-scale fields 
at different $Rm$ and $Pm$ are nevertheless likely to strongly affect the broader turbulent dynamics and how magnetic fields and 
differential rotation interact. This particular issue cannot be easily examined within my simplified framework.

\section{Conclusions and discussion}
Using a parameter scan and analysis of maximally simplified three-dimensional cartesian MHD simulations of
large-scale non-linear helical dynamos with hemispheric distributions of turbulent kinetic helicity, 
I have provided detailed numerical evidence and phenomenological
arguments for the existence of an asymptotic ultimate non-linear regime of the large-scale solar dynamo  
involving magnetic helicity fluxes between hemispheres. I obtained corresponding numerical solutions 
at both $Pm>1$ and $Pm<1$, and put forward  physical interpretations of how the nature of this 
large-scale dynamo changes in the $Re$-$Rm$ parameter space. The results, together with the recent study of \cite{branden25} with shear
and rotation, suggest that these ultimate solutions can currently only be obtained in simplified numerical DNS set-ups where most
of the numerical resolution can be put in resolving turbulent structures and transport processes.
The large scatter between global solar dynamo simulations outcomes has long been puzzling modellers \citep{charbonneau20}.
Our analysis highlights that they currently populate a non-asymptotic regime
in parameter space where results are highly sensitive to $Re$ and $Rm$, and likely more generally to the specific
implementation of dissipative processes. 

Despite the underlying simplifying modelling assumptions required to achieve large $Rm$, the core 
physics driving this dynamo (and that of \cite{delsordo13} and \cite{branden25} for 
different flow forcings), is in essence the same as that available in
standard global solar dynamo models, in which hemispheric 
distributions of kinetic helicity have clearly been diagnosed \citep{simard16,strugarek18}. 
Since a generic $\alpha^2$-type mechanism and equally generic turbulent fluxes drive these
solutions in the ultimate regime, there is good reason to believe that they bear some relevance to the Sun.  
However, this simplified approach has its own limitations. \cite{branden25} have recently taken on 
looking at the explicit role of shear and rotation using a similar approach. Their results
suggest that the large-scale field does not asymptote towards a small-value in this case.
A realistic interplay between the tachocline, differential rotation, and this dynamo is yet to be studied
though, and so are the geometric interplay between convection and the Coriolis force to inject helicity, 
meridional circulation, stratification, and magnetic buoyancy effects, which are all likely to affect 
the overall picture.

The high resolution and estimated $Rm=u_\mathrm{rms}/(k_f\eta)>5000$  required to enter 
the astrophysically relevant ultimate regime for solar-like scale separations between the mean field and turbulent injection scales 
are likely to hard-prevent global models with realistic geometries from approaching it in the foreseeable future,
at any reasonable computational cost.
It would be extremely useful now to produce diagnostics for these simulations,
such as those shown in \fig{fig4}, to gauge their lack of asymptoticity in $Rm$. 
In parallel, it might be possible to tune them into convection regimes with larger
injection scales, minimising the scale separation with the desired large-scale field to reach
the ultimate regime at lower $Rm_{I-U}$. Because magnetic-helicity fluxes play a key role in this regime, 
they notably provide the strong hemispheric coupling lacking in current global 3D models \citep{charbonneau20}. 
Hence, such a trade-off could produce more reliable solar-like dynamo cycles.
Another avenue would be to devise transport closures, for example based  on machine-learning informed
by the turbulent transport effects isolated here, as such techniques require prior 
physical insights into the relevant processes to be informative \citep[e.g.][]{ross23,eyring24}. 
Finally, a hydrodynamic anisotropic kinetic alpha effect (AKA/$\Lambda$)
similar to the large-scale helical MHD dynamo 
is also possible in helical flows \citep{frisch87}.  It could be interesting 
to investigate by similar means whether an ultimate large-$Re$ regime exists for this effect, and how
current helicity feeds back, in the saturated regime, on the helical flow driving the dynamo.

Much remains to be done to understand how the essence of the ultimate
regime distilled here and in \cite{branden25} translates to global models. Judging by the pace
at which the size of simulations has increased in recent years,
my concern is that this may require a full power plant, something we should be wary 
of avoiding in the current environmental emergency.

\begin{acknowledgements}
This work was granted access to the HPC resources of IDRIS under GENCI allocation
2020-A0080411406, and of CALMIP under allocation P09112.
2.5 MCPU hours were used, for an estimated 8.75 CO2eq tons
carbon footprint, using typical 2020 French supercomputer emission numbers \citep{berthoud20}.

This article is dedicated to my close friend and colleague, Nuno F. Loureiro, 
who tragically passed away in December 2025.
\end{acknowledgements}
\bibliographystyle{aa}
\bibliography{dynamo}

\appendix
\section{Magnetohydrodynamic model\label{appequations}}
The code solves the equations of incompressible non-linear magnetohydrodynamics,
namely the momentum (Navier-Stokes) equation
\begin{equation}
\label{eq:momincompressible}
\dpart{\vU}{t}+\vU\cdot\grad{\,\vU}=-\grad{\Pi}+\vB\cdot\grad\,\vB+\nu\Delta{\vU}+\vec{f}(\vec{x},t)~,
\end{equation}
and the induction equation
\begin{equation}
  \label{eq:inducincompressible}
\dpart{\vB}{t}+\vU\cdot\grad\,\vB=\vB\cdot\grad{\,\vU}+\eta\Delta\vB~,
\end{equation}
supplemented with
\begin{equation}
\div{\vU}=0~,\quad \div{\vB}=0~.
\end{equation}
Here $\vU$ and $\vB$ are the velocity and magnetic fields respectively (there is no mean flow, so a lower case variable is used for the former), 
$\vB$ is expressed as an Alf\'ven velocity, $\Pi=P+B^2/2$ is the total pressure and, as in R21, the forcing term is defined as
\begin{equation}
  \label{force}
  \begin{array}{l}
    \vec{f}(\vec{x},t)=k_f\,A_f\,\times \smallskip\\
\left(
    \begin{array}{c}
      \displaystyle{-2\sin\left(\f{2\pi y}{L_f}+\sin\omega_f t\right)\sin\f{2\pi z}{L_z}} \\
      \displaystyle{-2\cos\left(\f{2\pi x}{L_f}+\cos\omega_f t\right)\sin\f{2\pi z}{L_z}} \\
      \displaystyle{\sin\left(\f{2\pi x}{L_f}+\cos\omega_f
      t\right)+\cos\left(\f{2\pi y}{L_f}+\sin\omega_f t\right)}
      \end{array}
    \right)~, \end{array}
          \end{equation}
where $\omega_f$ and $A_f$ are a forcing frequency and amplitude,
 $L_x=L_y\equiv L_f$
 and $k_f=2\pi/L_f$ is the forcing wavenumber ($L_f=1$, $L_z=4L_f$, $\omega_f=1$
 and $A_f=0.1$ in all simulations).

\section{Simulation runs}
\begin{table*}[!htb]
\caption{Index of runs: V (viscous), M (moderate), T (turb.) runs 
from R21 ; S01 (Super-turb.) is new.
  $Re=u_\mathrm{rms}/(k_f\nu)$ and $Rm=u_\mathrm{rms}/(k_f\eta)$.\label{tab1}}
\centering\begin{tabular}{lccllrrrccc}
  \hline\hline  
  Run & $(N_x,N_y,N_z)$ & $L_z/L_f$ & $\nu^{-1}$ & $\eta^{-1}$ & $Pm$ & $Re$ & $Rm$ & $u_\mathrm{rms}$ & $B_\mathrm{rms}$ & $\meanB_\mathrm{rms}$\\\hline
  V02  & $(64,64,256)$    & 4 & 500 & 500 & 1 & 46.2 & 46.2 & 0.58 & 0.57 & 0.42 \\
  V03  & $(128,128,512)$  & 4 & 500 & 2000 & 4 & 39.9 & 159.7 & 0.50 & 0.54 & 0.21 \\
  V04  & $(128,128,512)$  & 4 & 500 & 8000 & 16 & 35.6 & 570.0 & 0.44 & 0.58 & 0.14  \\
  V05  & $(256,256,1024)$ & 4 & 500 & 16000 & 32 & 32.9 & 1054.0 & 0.41 & 0.59 & 0.12 \\\hline
  M02  & $(64,64,256)$    & 4 & 2000 & 500 & 0.25 & 197.6 & 49.4 & 0.62 & 0.56 & 0.38 \\
  M03  & $(128,128,512)$  & 4 & 2000 & 2000 & 1 & 177.0 & 177.0 & 0.56 & 0.59 & 0.32 \\
  M04  & $(128,128,512)$  & 4 & 2000 & 8000 & 4 & 163.3 & 653.2 & 0.51 & 0.60 & 0.18  \\
  M05  & $(256,256,1024)$ & 4 & 2000 & 16000 & 8 & 165.9 & 1327.5 & 0.52 & 0.60 & 0.12 \\\hline
  T02  & $(128,128,512)$  & 4 & 8000 & 500 & 0.062 & 846.6 & 52.9 & 0. 66 & 0.58 & 0.42 \\
  T03  & $(128,128,512)$  & 4 & 8000 & 2000 & 0.25 & 749.3 & 187.3 & 0.59 & 0.58 & 0.25 \\
  T04  & $(128,128,512)$  & 4 & 8000 & 8000 & 1 & 748.1 & 748.1 & 0.58 & 0.60 & 0.16 \\
  T05  & $(256,256,1024)$ & 4 & 8000 & 16000 & 2 & 683.4 & 1366.8 & 0.54 & 0.63 & 0.17 \\
  T06  & $(512,512,2048)$ & 4 & 8000 & 32000 & 4 & 694.5 & 2778.0 & 0.55 & 0.62 & 0.12 \\\hline
  S01  & $(512,512,2048)$ & 4 & 32000 & 16000 & 0.5 & 2904.5 & 1452.2  & 0.57 & 0.63 & 0.15 \\\hline
  
  \hline\hline
\end{tabular}
\end{table*}

\begin{figure}[!htbp]
\centering\includegraphics[width=\columnwidth]{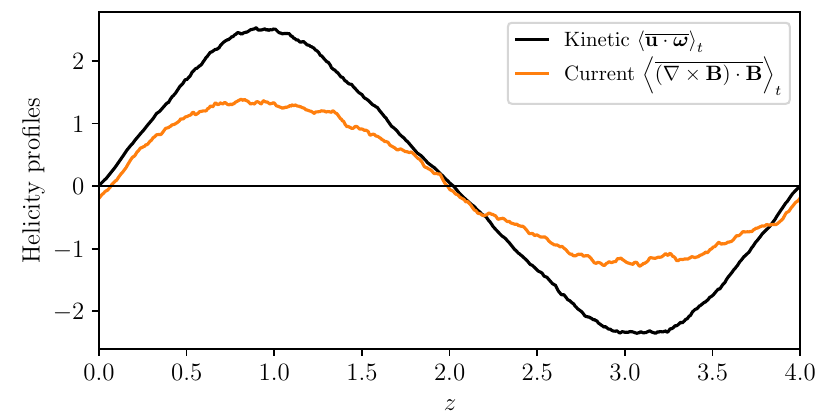}
\caption{Time-averaged and $(x,y)$-averaged kinetic and current helicity $z$-profiles in run S01
  ($Rm\simeq 1400$, $Re\simeq 2900$).\label{fig2}}
\end{figure}

\begin{figure}[!htbp]
\centering\includegraphics[width=\columnwidth]{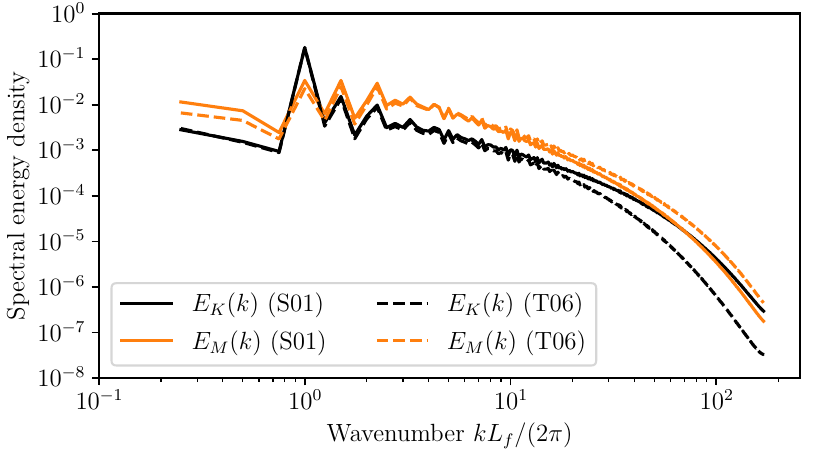}
\caption{Energy spectra for run S01
    ($Rm\simeq 1450$, $Re\simeq 2900$, full line) and run T06 ($Rm
\simeq 2800$, $Re\simeq 700$, dashed line).\label{fig3}}
\end{figure}

\section{Three regimes\label{regimesection}}
As in R21, I start with the magnetic-helicity budget
\begin{equation}
\label{budget}
  \dpart{}{t}(\vA\cdot\vB)+\div{\vec{F}_{\mathcal{H}_m}}=-2\eta\,{\vJ}\cdot\vB~,
\end{equation}
where $\vec{F}_{\mathcal{H}_m} = c\left(\varphi\vB+\vE\times\vA\right)$
is the magnetic-helicity flux, $c$ is the speed of light,
$\vE$ is the electric field, $\varphi$ is the electrostatic
potential, $\vA$ is the magnetic vector potential, and $\vJ=\curl{\vB}$ is the electric current
(in all that follows, I work in the Coulomb gauge). Decomposing $\vB$ into its mean (average 
over the $(x,y)$ plane) and fluctuations, $\vB(\vec{r},t)=\meanvB(z,t)+\fluctvB(\vec{r},t)$ 
(and similarly for $\vJ$ and $\vA$), \equ{budget} is separated into large-scale/mean 
and small-scale/fluctuating parts,
\begin{eqnarray}
\label{LShelicity}
\dpart{}{t}(\meanvA\cdot\meanvB)+\div{\mean{\vec{F}}_{\mathcal{H}_{m,\mathrm{mean}}}} & = & 2\,\meanvEMF\cdot\meanvB-2\eta\left(\curl{\meanvB}\right)\cdot\meanvB~,\\
\label{SShelicity}
\dpart{}{t}(\mean{\fluctvA\cdot\fluctvB})+\div{\mean{\vec{F}}_{\mathcal{H}_{m,\mathrm{fluct}}}} & = & -2\,\meanvEMF\cdot\meanvB-2\eta\,\mean{(\curl{\fluctvB})\cdot\fluctvB}~,
\end{eqnarray}
where
$\mean{\vec{F}}_{\mathcal{H}_{m,\mathrm{mean}}}$ and $\mean{\vec{F}}_{\mathcal{H}_{m,\mathrm{fluct}}}$ denote the mean flux of large and 
small-scale magnetic helicity respectively (see R21 for their detailed definition).  $\vEMF=\fluctvU\times\vB$ is the electromotive force 
(EMF) for a flow $\fluctvU$, and we have used $\vEMF\cdot\vB=0$, so that the EMF itself only redistributes helicity 
into large and small-scale parts.
In a statistically steady non-linear state, \equs{LShelicity}{SShelicity} reduce to
\begin{equation}
\label{fraction}
\f{\left|\left<\div{\mean{\vec{F}}_{\mathcal{H}_{m,\mathrm{fluct}}}}\right>+2\eta\,\left<\overline{\vec{j}\cdot \vec{b}}\right>\right|}{\left|\left<\div{\mean{\vec{F}}_{\mathcal{H}_{m,\mathrm{mean}}}}\right>+2\eta\,\left<\overline{\vec{J}}\cdot\overline{\vec{B}}\right>\right|}=1~.
\end{equation}

\begin{figure}[h]
\centering\includegraphics[width=\columnwidth]{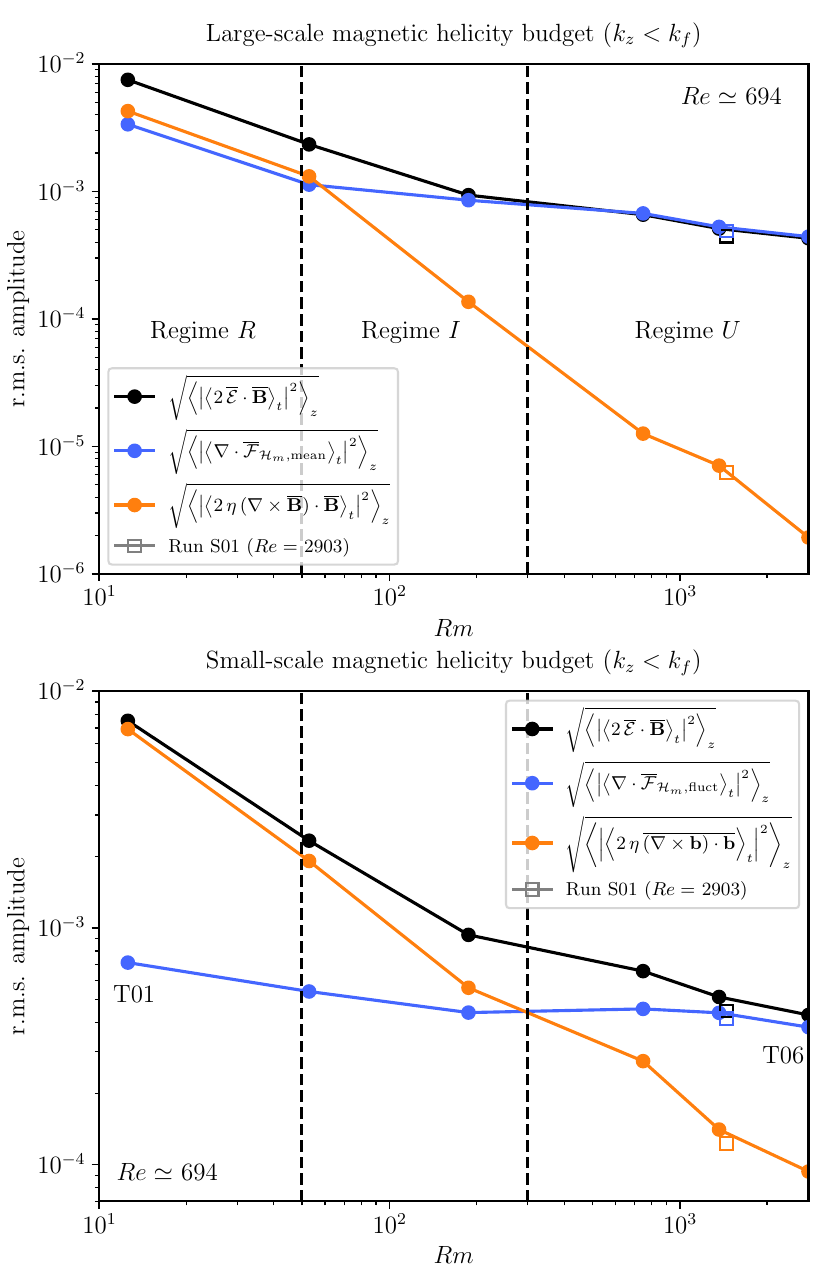}
\caption{Helicity budgets for the T and S runs as a function of $Rm$, with corresponding regime tags and qualitative separations between regimes 
(vertical dashed lines). The lines and full circles correspond to the T runs ($Re\simeq 694$ based on T06) in R21; the empty squares show the 
same quantities for the new $Re\simeq 2900$, $Pm=0.5$ run S01.\label{fig4}}
\end{figure}

\Equ{fraction} gives rises to three distinct regimes, established in R21 and now explicitly tagged here in \fig{fig4}:
\begin{enumerate}
\item A resistively dominated ($R$) low-$Rm$ regime, where the dominant balance in both \equs{LShelicity}{SShelicity} is between the EMF and resistive terms, 
so that the resistive terms dominate both the numerator and denominator of \equ{fraction}. The same low-$Rm$ helicity budget dominant balance
was previously obtained by \cite{mitra10b} using a similar, albeit distinct forcing.
\item An intermediate-$Rm$ ($I$) regime, where the resistive term still balances the EMF in \equ{SShelicity}, and thefore still dominates the numerator in \equ{fraction}, but the dominant balance in \equ{LShelicity} is now between the EMF and helicity flux divergence term, so that the mean flux of large-scale helicity dominates in the denominator of \equ{fraction}. This regime was also found, and first probed by \cite{delsordo13} using a similar, albeit distinct forcing.
\item A large $Rm$, asymptotic ultimate ($U$) regime, where the resistive helicity dissipation terms are subdominant in both \equs{LShelicity}{SShelicity}, and the mean fluxes of small-scale and large-scale helicities consequently dominate the numerator and denominator in \equ{fraction}. This regime was first probed by runs T05-T06 in R21 at $Pm>1$, but \fig{fig4} further shows that the new run S01 at $Pm=0.5$ and $Re=2900$ also lies in this regime, and that its helicity budgets are almost identical to those obtained for run T05 at the same $Rm$ but lower $Re$.
\end{enumerate}
Let us denote the transition $Rm$ between the $R$ and $I$ regimes as $Rm_{R-I}$, and that between the intermediate and ultimate regimes as $Rm_{I-U}$.
The latter is the likely regime of astrophysical interest, and a key question follows: how are current global solar dynamo simulations 
positioned with respect to $Rm_{I-U}$ ? This question can be addressed using a phenomenlogical diffusive flux model, described below in Appendix~\ref{diffusive}.

\section{Diffusive helicity-flux model\label{diffusive}}
\begin{figure}
\includegraphics[width=\columnwidth]{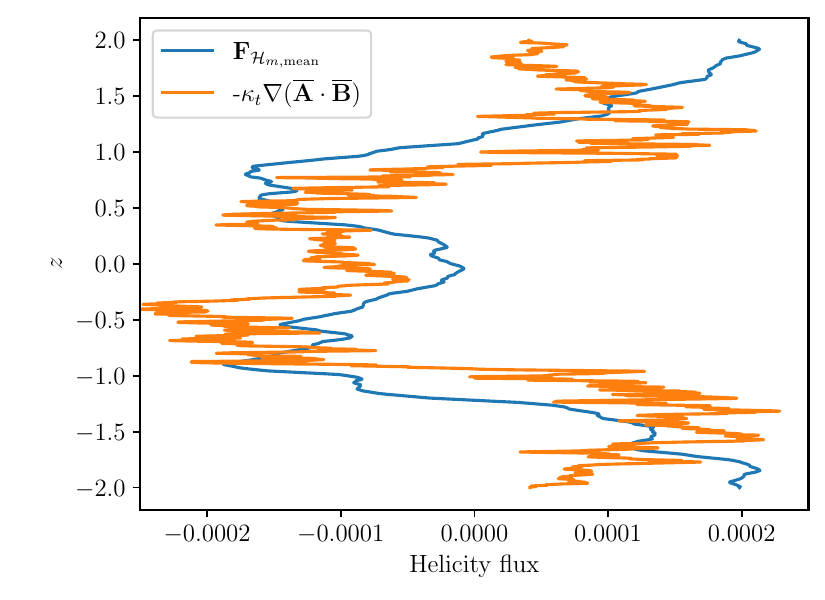}
\includegraphics[width=\columnwidth]{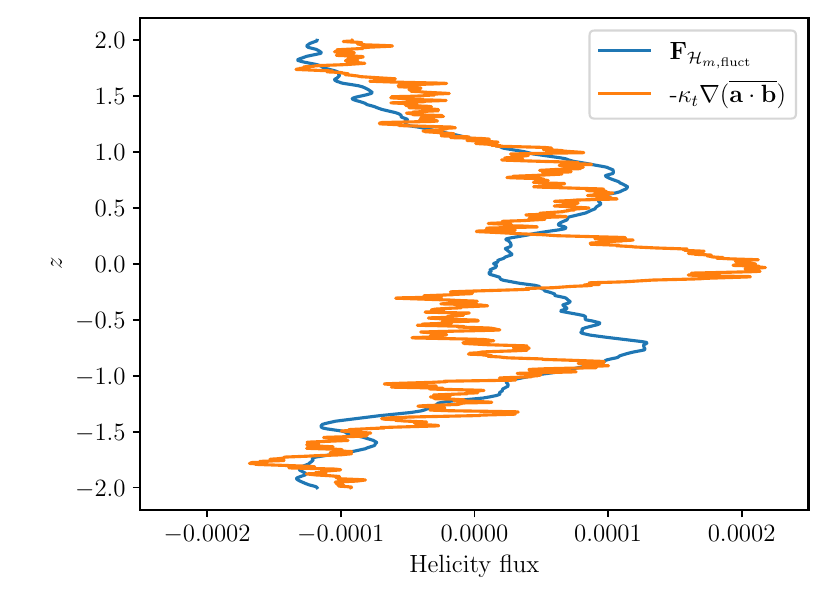}
\caption{Turbulent fluxes of large and small-scale magnetic helicity in the T06 run at $Rm=2800$, and their respective diffusive fits. The best fit for the turbulent flux of large-scale helicity gives $\kappa_t=0.6 \,u_\mathrm{rms}/(3\,k_f)$, and that for the flux or small-scale helicity $\kappa_t=0.55 \,u_\mathrm{rms}/(3\,k_f)$.\label{diffusivefit}}
\end{figure}

To make phenomenological progress on the problem, I follow \cite{branden01d,mitra10b} by devising the folllowing diffusive helicity flux model postulating
\begin{eqnarray}
\label{LSflux}
\mean{\vec{F}}_{\mathcal{H}_{m,\mathrm{mean}}} & \hat{=} & -\kappa_t\grad{\left(\overline{\vec{A}}\cdot\overline{\vec{B}}\right)}~,\\
\label{SSflux}
\mean{\vec{F}}_{\mathcal{H}_{m,\mathrm{fluct}}} & \hat{=} & -\kappa_t\grad{\left(\overline{\vec{a}\cdot\vec{b}}\right)}~.
\end{eqnarray}
Using a dimensional mixing-length argument, we expect
\begin{equation}
\kappa_t=\xi\f{u_\mathrm{rms}}{3\,k_f},
\end{equation}
where $\xi$ is a numerical prefactor to be determined. An effective two-scale parametrisation is adopted 
by introducing $\overline{k}$, which stands for the dominant wavenumber of the large-scale magnetic field, 
and $k_\mathrm{eff}\geq k_f$, an effective wavenumber of turbulent magnetic fluctuations. Also introducing
$\overline{\theta}=\overline{k}\,\meanvA\cdot\meanvB/\meanB^2$ and $\theta_\mathrm{eff}=k_\mathrm{eff}\,\mean{\fluctvA\cdot\fluctvB}/\mean{\fluctB^2}$, 
the fractional helicities of the large-scale and fluctuation magnetic field respectively, we can express the different terms in \equ{fraction} as
\begin{eqnarray}
\div{\mean{\vec{F}}_{\mathcal{H}_{m,\mathrm{mean}}}}& \sim &\overline{\theta}\,\xi\f{ u_\mathrm{rms}}{3} \f{\overline{k}}{k_f}\overline{B}^2~,\\
\div{\mean{\vec{F}}_{\mathcal{H}_{m,\mathrm{fluct}}}} & \sim & \theta_\mathrm{eff}\,\xi \f{u_\mathrm{rms}}{3} \left(\f{\overline{k}^2}{k_\mathrm{eff}k_f}\right)\overline{b^2}~,
\end{eqnarray}
\begin{eqnarray}
2\eta \,\overline{\vec{J}}\cdot{\overline{\vec{B}}} & \sim & 2\eta\,\overline{\theta}\,\overline{k}\,\overline{B}^2~,\\
2\eta\,\overline{\vec{j}}\cdot{\overline{\vec{b}}} & \sim & 2\eta\, \theta_\mathrm{eff} k_\mathrm{eff}\,\overline{b^2}~.
\end{eqnarray}
\Equ{fraction} then provides a simple prescription for the saturation level of the large-scale field,
\begin{equation}
\f{\left<\overline{B}^2\right>}{\left<\overline{b^2}\right>}= \left(\frac{\theta_\mathrm{eff}}{\overline{\theta}}\right)\,\left(\f{\overline{k}}{k_\mathrm{eff}}\right)\,\f{1+ Rm_{\tiny 2-3}/Rm}{1+Rm_{\tiny 1-2}/Rm}~,
\end{equation}
with
\begin{eqnarray}
Rm_{R-I} & = & \frac{6}{\xi}~,\\
Rm_{I-U} & = & \frac{6}{\xi}\left(\f{k_\mathrm{eff}}{\overline{k}}\right)^2~.
\label{Rm23}
\end{eqnarray}
In the low-$Rm$ limit (resistive regime),
\begin{equation}
\label{bbar1}
\frac{\left<\overline{B}^2\right>}{\left<\overline{b^2}\right>} = \left(\f{\theta_\mathrm{eff}}{\overline{\theta}}\right)\,\left(\f{k_\mathrm{eff}}{\overline{k}}\right)~.
\end{equation}
In the intermediate-$Rm$ regime,
\begin{equation}
\label{bbar2}
\frac{\left<\overline{B}^2\right>}{\left<\overline{b^2}\right>} = \frac{6}{\xi}\left(\f{\theta_\mathrm{eff}}{\overline{\theta}}\right)\,\left(\f{k_\mathrm{eff}}{\overline{k}}\right)\frac{1}{Rm}~.
\end{equation}
Finally, in the large-$Rm$ limit (ultimate regime),
\begin{equation}
\label{bbar3}
\frac{\left<\overline{B}^2\right>}{\left<\overline{b^2}\right>} = \left(\f{\theta_\mathrm{eff}}{\overline{\theta}}\right)\,\left(\f{\overline{k}}{k_\mathrm{eff}}\right)~.
\end{equation}
We are now finally in a position to calibrate the above model against the suite of simulations.
\Fig{diffusivefit} for run T06 first shows that the diffusion approximation, \equs{LSflux}{SSflux}, works reasonably well to model both fluxes of large and small-scale magnetic helicity in the highest-$Rm$ simulations, giving consistent values $\xi\simeq 0.55-0.6$. This value is also consistent with previous results obtained by \cite{delsordo13} using a different turbulent forcing ($\xi$ here is the same as $\kappa_\mathrm{f}/\eta_\mathrm{t}$ in their Table~2).

\begin{figure}[!hbp]
\includegraphics[width=\columnwidth]{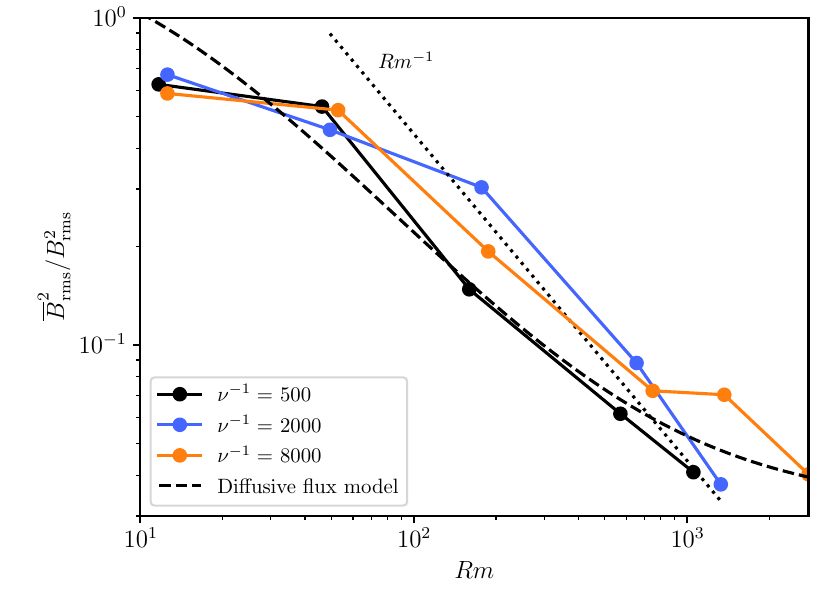}
\caption{Comparison of simulations with the diffusive-flux model for $\left<\overline{B}^2\right>/\left<\overline{b^2}\right>$ as a function of $Rm$. 
The model parameters, extracted from fits to the simulations, are $\xi\sim 0.55$, $\overline{k}=3$, $k_\mathrm{eff}/\overline{k}=7.8$, 
$\theta_\mathrm{eff}=0.15$, $\overline{\theta}=0.6$, corresponding to $Rm_{R-I}=11$ and $Rm_{I-U}=670$. \label{fit}}
\end{figure}

\Fig{fit} shows that this phenomenological effective model can provide a good fit to the results of the simulations, and a sound basis for their interpretation.
Two key observations are in order. First, both the model and numerical results suggest a sluggish asymptote of this large-scale dynamo towards the large-$Rm$ limit, that even run T06 at $Rm=2800$, corresponding to the rightmost point in the figure, barely starts to trace. Second, as already pointed out by \cite{branden01d,mitra10b} and in the main text, the transition $Rm_{I-U}$ to the ultimate large-$Rm$ regime scales as the square of the scale separation between the large-scale field and turbulent forcing scale, see \equ{Rm23} with $k_\mathrm{eff}\sim k_f$. For this set-up, \fig{fit} shows that $Rm_{I-U}\simeq 670$, with $k_\mathrm{eff}/\overline{k}=7$ consistent with $k_f L_z/(2\pi)=4$. Were the scale separation between the large-scale dynamo field and turbulent forcing scale significantly larger than that, as expected in global spherical models of the solar dynamo, that transition is expected to occur at much higher $Rm_{I-U}$ than in the present carefully optimised set-up.

\end{document}